\documentclass[pra,showpacs,twocolumn,nofootinbib]{revtex4}
\usepackage{graphicx}
\usepackage{bm,amsmath}
\usepackage{dcolumn}
\usepackage{natbib}
\usepackage
{hyperref}

\newcommand{\sref}[1]{Sec. \ref{#1}}
\newcommand{\Eref}[1]{Eq.~(\ref{#1})}
\newcommand{\tref}[1]{Table~\ref{#1}}

\begin{document}

\title{$\Lambda$-doublet spectra of diatomic radicals and
their dependence on fundamental constants}

\author{M. G. Kozlov}
\affiliation{Petersburg Nuclear Physics Institute, Gatchina 188300,
             Russia}
\date{
\today}

\begin{abstract}
$\Lambda$-doublet spectra of light diatomic radicals have high
sensitivity to the possible variations of the fine structure
constant $\alpha$ and electron-to-proton mass ratio $\beta$. For
molecules OH and CH sensitivity is further enhanced because of the
$J$-dependent decoupling of the electron spin from the molecular
axis, where $J$ is total angular momentum of the molecule. When
$\Lambda$-splitting has different signs in two limiting coupling
cases $a$ and $b$, decoupling of the spin leads to the change of
sign of the splitting and to the growth of the dimensionless
sensitivity coefficients. For example, sensitivity coefficients for
the $\Lambda$-doublet lines $J=\tfrac92$ of the $\Pi_{1/2}$ state of
OH molecule are on the order of $10^3$.
\end{abstract}

\maketitle

\section{Introduction}

At present an intensive search is going on for the possible space
and time variations of fundamental constants (FC). On a short time
scale very tight bounds on such variations were obtained in
laboratory experiments \cite{Ros08,SBC08}. On the other hand
astrophysical observations provide information on the variation of
FC on the timescale of the order of $10^{10}$ years. Here results of
Ref.\ \cite{MWF03} indicate variation on the level of five sigma,
 $\Delta\alpha/\alpha= (-0.57 \pm 0.11)\times 10^{-5}$.
At the same time, Ref.\ \cite{SCP04} reports no variation,
 $\Delta\alpha/\alpha= (-0.06\pm 0.06)\times 10^{-5}$,
and Ref.\ \cite{LML07} reports variation of the opposite sign,
 $\Delta\alpha/\alpha= (+0.54\pm 0.25)\times 10^{-5}$. An intermediate
timescale $\sim 2\times 10^9$ years is tested by the Oklo phenomenon
\cite{Shl76,FW09}.

Recently there was much attention in this context to the microwave
spectra of molecules. Generally these spectra are sensitive to
possible variations of the electron to proton mass ratio
$\beta=m_e/m_p$. When fine and hyperfine structures are involved,
they also become sensitive to variations of the fine structure
constant $\alpha$ and nuclear $g$-factor $g_\mathrm{nuc}$. There
were several proposals of microwave experiments with diatomic
molecules. Rotational microwave spectra were used numerous times to
study time variation of fundamental constants in astrophysics.
However, all such lines have same dependence on FC,
$\delta\omega_\mathrm{rot}/\omega_\mathrm{rot} = \delta\beta/\beta$,
so one needs to use reference lines with different dependence on FC.
In the microwave band there are several examples of such lines (see
\tref{tab_main_lines}).

\begin{table*}[bht]
\caption{Quantum numbers and frequencies of microwave lines used for
astrophysical studies of possible variation of FC. For $\Lambda$
transitions of CH and OH molecules only one of the strongest
hyperfine components is given. Ammonia inversion transition has
rotational structure described by quantum numbers $J$ and $K$, where
$K$ is projection of the angular momentum $\bm{J}$ on the molecular
axis, and smaller hyperfine structure, described by the total
angular momentum quantum number $F$ and intermediate quantum number
$F_1$. Here we present one of the 18 hyperfine components of the
inversion line $(J,K)=(1,1)$.}

\label{tab_main_lines}

\begin{tabular}{cdldr}
 \hline\hline\\[-7pt]
 \multicolumn{1}{c}{Atom/molecule}
 &\multicolumn{1}{c}{$\lambda$ (cm)}
 &\multicolumn{1}{c}{Quantum numbers}
 &\multicolumn{1}{c}{Frequency (MHz)}
 &\multicolumn{1}{c}{Ref.}
\\
\hline
 H   & 21  & $1s_{1/2},\, F=0-1$              & 1420.405751767(1) & \cite{EDB71}  \\
 OH  & 18  & $\Pi_{3/2},\,J=\tfrac32,  F=2$   & 1667.358996(4) & \cite{HLS06}   \\[1mm]
 CH  &  9.0& $\Pi_{1/2},\,J=\tfrac12,  F=1$   & 3335.481(1)    & \cite{MMB06}   \\[1mm]
 CH  & 42  & $\Pi_{3/2},\,J=\tfrac32,  F=2$   &  701.677(10)   & \cite{ZT85}    \\[1mm]
 CH  &  4.1& $\Pi_{1/2},\,J=\tfrac32,  F=2$   & 7348.419(1)    & \cite{MMB06}   \\[1mm]
 NH$_3$ & 1.2
           & $(J,K)=(1,1);\,F,F_1=\tfrac12,1$ & 23694.4591(1)  & \cite{Kuk67} \\
\hline\hline
\end{tabular}
\end{table*}

First, the famous 21 cm hydrogen hyperfine line depends on all three
FC, $\delta\omega_\mathrm{hf}/\omega_\mathrm{hf} = \delta{\cal
F}/{\cal F}$, ${\cal F}=\alpha^2\beta g_\mathrm{nuc}$ (note that the
21 cm line of hydrogen was used to constrain variation of FC as
early as 1956 \cite{Sav56}). Second, the 18 cm $\Lambda$-doublet
line of OH molecule depends on $\alpha$ and $\beta$ as follows:
$\delta\omega_\mathrm{OH}/\omega_\mathrm{OH} = \delta{\cal F}/{\cal
F}$, ${\cal F}=\alpha^{-1.14}\beta^{2.57}$ \cite{Dar03,CK03,KCG04}.
Third, the 1.2 cm inversion line of ammonia depends only on $\beta$,
$\delta\omega_\mathrm{inv}/\omega_\mathrm{inv} =
4.46\,\delta\beta/\beta$ \cite{VKB04,FK07a}. Finally, the fine
structure far infrared 158 $\mu$m line of C~II is sensitive only to
$\alpha$, $\delta\omega_\mathrm{fs}/\omega_\mathrm{fs} =
2\delta\alpha/\alpha$. All these four reference lines were used in
combination with some rotational lines to put strong limits on
variation of FC \cite{MWF01c,KCL05,FK07a,MFMH08,LRK08}.

If the Hydrogen 21-cm line is used as a reference for 18-cm OH line, the
combination of constants, which is constrained, has the form \cite{KCL05}:
 \begin{align}\label{tildeF}
 {\cal F}=\alpha^{3.14}\beta^{-1.57} g_\mathrm{nuc}\,.
 \end{align}
The most tight limit on the variation of $\cal F$ was obtained from
observations of the absorber at the redshift $z=0.765$ and the
$z=0.685$ gravitational lens, Ref.~\cite{KCL05}:
 \begin{align}\label{OH1}
 \delta {\cal F}/{\cal F} = \left(0.44 \pm0.36^\mathrm{stat}\pm
 1.0^\mathrm{syst}\right)\times 10^{-5}\,.
 \end{align}

For OH molecule at least two more $\Lambda$-doublet lines were
detected from interstellar medium in addition to the lowest 18 cm
line, which was used in Ref.~\cite{KCL05}. Sensitivity coefficient
for these lines were found in Ref.~\cite{KC04}. They appeared to be
rather different from those of the lowest $\Lambda$-doublet line.
Therefore, it is possible to use different $\Lambda$-doublet lines
of the OH molecule to place a limit on the variation of fundamental
constants without using reference lines of other species. This can
help to eliminate systematic effect from the different velocity
distributions of different species in molecular clouds. Two lowest
$\Lambda$-doublet lines of CH molecule (9 cm and 42 cm) were
detected in the interstellar medium \cite{RES74,ZT85}. Recently
Christian Henkel and Karl Menten suggested that these lines can be
used for astrophysical search of the time-variation of fundamental
constants \cite{HM08}. There are also several other light molecules
with $\Lambda$-doubling, where microwave spectra were observed in
the interstellar medium. For example, first extragalactic microwave
rotational spectra of NO were observed in \cite{MMM03}. Therefore,
we decided to study sensitivities of the $\Lambda$-doublet lines to
the variation of the fundamental constants in a more systematic way.

Astrophysical studies of variation of fundamental constants require
accurate knowledge of the laboratory frequencies. In the microwave
band it is not so rare that the accuracy of the astrophysical
observations is higher than the accuracy of the respective
laboratory measurements. Therefore, some of the recommended
``laboratory'' frequencies are actually recalculated from
astrophysical spectra (see, for example, \cite{ZT85,PDD09}). This
method is based on the assumption that different lines from the same
distant object have the same redshifts. Thus, the redshift is first
determined from one set of lines and then is used to find rest frame
frequencies of the other set of observed lines. The logic in these
works is opposite to the one used in the search of the variation of
FC. In such a search one looks for the difference in the
\textit{apparent} redshifts of the lines from the same object and
compare these differences to the sensitivities of respective lines
to variation of the constants to get information on constant
variation.

Recently the laboratory frequencies of all four hyperfine components
of the 18~cm line of OH molecule were measured with a record
precision $(< 10^{-9})$ \cite{LMH06,HLS06}. Also, the frequencies of
all three components of the 9 cm $\Lambda$-doublet line $J=\tfrac12$
in CH molecule were recently remeasured in Ref.~\cite{MMB06} with
the accuracy of 0.1 ppm, or better ($1\,\mathrm{ppm} = 10^{-6}$).
This opens possibility to study variation of fundamental constants
at the level below 1~ppm. Such studies can supplement the limits on
$\beta$-variation based on the observations of the ammonia inversion
line \cite{FK07a,MFMH08,LMK08,KMM09} because $\Lambda$-doublet lines
are sensitive to variation of $\alpha$ and $\beta$, while ammonia
line is sensitive only to $\beta$. Moreover, as we will show below,
because of the decoupling of the electron spin from the molecular
axis, the sensitivity coefficients here strongly depend on the
rotational quantum numbers. Therefore, if more than one line is
observed, it may be possible to obtain model independent limits on
variation of both constants. Sensitivity to the third constant
$g_\mathrm{nuc}$ is typically much weaker, except for some low
frequency lines where hyperfine contribution to transition frequency
becomes significant. If such lines are observed, it is possible to
make \textit{full} experiment and constrain variation of \textit{all
three constants.}

Additional motivation to the present work comes from rapid progress
in laboratory experiments with cold and ultracold molecules. New
laboratory techniques can make it possible to use molecular
$\Lambda$-doublet lines for laboratory tests on variation of FC. The
most recent developments in this field are summarized in the review
\cite{CDKY09}.

In this paper we estimate sensitivity coefficients of different
$\Lambda$-doublet lines to variations of constants $\alpha$,
$\beta$, and $g_\mathrm{nuc}$. The analysis is basically the same
for all light molecules in the $^2\Pi_{1/2}$, or $^2\Pi_{3/2}$
states. We include several of them here, for which there is
sufficient data in the databases of microwave molecular spectra
\cite{NIST_diatomics,JPL_Catalog,CDMS}. We use this data to find
parameters of the effective spin-rotational Hamiltonian and to
calculate sensitivity coefficients.

\section{Sensitivity coefficients}\label{sensitivity}

We restrict ourselves to the case of the diatomic radicals in
doublet states $^2\Pi_{1/2}$, or $^2\Pi_{3/2}$. Let us define
dimensionless sensitivity coefficients to the variation of FC so
that:
 \begin{align}\label{K-factors}
 \frac{\delta\omega}{\omega}
 = K_\alpha\frac{\delta\alpha}{\alpha}
 + K_\beta\frac{\delta\beta}{\beta}
 + K_g\frac{\delta g_\mathrm{nuc}}{g_\mathrm{nuc}}\,.
 \end{align}

Dimensionless sensitivity coefficients $K_i$ are most relevant in
astrophysics, where lines are Doppler broadened, so
$\Gamma\approx\Gamma_D=\omega\times\Delta v/c$, where $\Delta v$ is
velocity variance and $c$ is the speed of light. The redshift of a
given line is defined as $z_i=\omega_{\mathrm{lab},i}/\omega_i-1$.
Frequency shift \eqref{K-factors} leads to the change in the
apparent redshifts of individual lines. The difference in the
redshifts of two lines is given by:
 \begin{align}\label{redshifts1}
 \frac{z_i-z_j}{1+z}
 = - \Delta K_\alpha\frac{\delta\alpha}{\alpha}
 - \Delta K_\beta\frac{\delta\beta}{\beta}
 - \Delta K_g\frac{\delta g_\mathrm{nuc}}{g_\mathrm{nuc}}\,.
 \end{align}
where $z$ is the average redshift of both lines and $\Delta
K_\alpha=K_{\alpha,i}-K_{\alpha,j}$, etc. We can rewrite
\Eref{redshifts1} in terms of the variation of a single parameter
$\cal F$:
 \begin{align}\label{redshifts2}
 \frac{z_i-z_j}{1+z}
 = - \frac{\delta{\cal F}}{\cal F}\,,
 \quad
 {\cal F}\equiv
 \alpha^{\Delta K_\alpha}
 \beta^{\Delta K_\beta}
 g_\mathrm{nuc}^{\Delta K_g}\,.
 \end{align}
The typical values of $\Delta v$ for the extragalactic spectra is on
the order of few km/s. This determines the accuracy of the redshift
measurements on the order of $\delta z=10^{-5}$ --~$10^{-6}$,
practically independent on the transition frequency. Therefore,
\textit{the sensitivity of astrophysical spectra to variations of FC
directly depend on $\Delta K_i$}.

In optical range sensitivity coefficients are typically on the order
of $10^{-2}$ --~$10^{-3}$, while in microwave and far infrared
frequency regions they are typically on the order of unity. In fact,
as we will see below, in some special cases sensitivity coefficients
can be much greater that unity. This makes observations in microwave
and far infrared wavelength regions potentially more sensitive to
variations of FC, as compared to observations in optical region.
Because of the lower sensitivity, systematic effects in optics may
be significantly larger (for the most recent discussion of the
systematic effects see \cite{GWW09} and references therein).

In \sref{analytic} we briefly recall the theory of $\Lambda$- and
$\Omega$-doubling in the pure coupling cases $a$ and $b$ and find
respective sensitivity coefficients. After that we will calculate
sensitivity coefficients for particular molecules using simplified
variant of effective Hamiltonian from Ref.\ \cite{MD72}. This
Hamiltonian accounts for decoupling phenomena and for the hyperfine
structure of $\Lambda$-doublets. We fit free parameters of this
Hamiltonian to match experimental frequencies. After that we use
numerical differentiation to find sensitivity coefficients.

\subsection{$\Lambda$-doubling and
$\Omega$-doubling}\label{analytic}

Consider electronic state with nonzero projection $\Lambda$ of the
orbital angular momentum on the molecular axis. The spin-orbit
interaction couples electron spin $\bm{S}$ to the molecular axis,
its projection being $\Sigma$. To a first approximation the
spin-orbit interaction is reduced to the form
$H_{so}=A\Lambda\Sigma$. Total electronic angular momentum
$\bm{J}_e=\bm{L}+\bm{S}$ has projection $\Omega$ on the axis,
$\Omega=\Lambda+\Sigma$. For a particular case of $\Lambda=1$ and
$S=\tfrac12$ we have two states $\Pi_{1/2}$ and $\Pi_{3/2}$ and the
energy difference between them is: $E(\Pi_{3/2})-E(\Pi_{1/2})=A$.

Rotational energy of the molecule is described by the Hamiltonian:
 \begin{subequations}\label{Hrot}
 \begin{align}
 \label{Hrot1}
  H_\mathrm{rot} &=B(\bm{J}-\bm{J}_e)^2\\
  \label{Hrot2}
  &=B\bm{J}^2-2B(\bm{JJ}_e)+B\bm{J}_e^2\,
 \end{align}
 \end{subequations}
where $B$ is rotational constant and $\bm{J}$ is the total angular
momentum of the molecule. The first term in expression \eqref{Hrot2}
describes conventional rotational spectrum. The last term is
constant for a given electronic state and can be added to the
electronic energy.\footnote{Note that this term contributes to the
separation between states $\Pi_{1/2}$ and $\Pi_{3/2}$. This becomes
particularly important for light molecules, where constant $A$ is
small.} The second term describes $\Omega$-doubling and is known as
Coriolis interaction $H_\mathrm{Cor}$.

If we neglect Coriolis interaction, the eigenvectors of Hamiltonian
\eqref{Hrot} have definite projections $M$ and $\Omega$ of the
molecular angular momentum $\bm{J}$ on the laboratory axis and on
the molecular axis respectively. In this approximation the states
$|J,M,\Lambda,\Sigma,\Omega\rangle$ and
$|J,M,-\Lambda,-\Sigma,-\Omega\rangle$ are degenerate,
$E_{J,\pm\Omega}=BJ(J+1)$. Coriolis interaction couples these states
and removes degeneracy. New eigenstates are the states of definite
parity $p=\pm 1$ \cite{BC03}:
 \begin{align}\label{patity_states}
 |J,M,\Omega,p\rangle
 &= \left(|J,M,\Omega\rangle
 +p(-1)^{J-S} |J,M,-\Omega\rangle\right)
 /\sqrt{2}\,,
 \end{align}
Operator $H_\mathrm{Cor}$ can only change quantum number $\Omega$ by
one, so the coupling of states $|\Omega\rangle$ and
$|-\Omega\rangle$ takes place in the $2\Omega$ order of the
perturbation theory in $H_\mathrm{Cor}$.

$\Omega$-doubling for the state $\Pi_{1/2}$ happens already in the
first order in Coriolis interaction, but has additional smallness
from the spin-orbit mixing. Operator $H_\mathrm{Cor}$ can not
directly mix degenerate states
$|\Lambda=1,\Sigma=-\tfrac12,\Omega=\tfrac12\rangle$ and
$|\Lambda=-1,\Sigma=\tfrac12,\Omega=-\tfrac12\rangle$ because it
requires changing $\Lambda$ by two. Therefore, we need to consider
spin-orbit mixing between $\Pi$ and $\Sigma$ states:
 \begin{align}\label{so-mixing}
 |\Omega=\tfrac12\rangle
 &=|\Lambda=1,\Sigma=-\tfrac12,\Omega=\tfrac12\rangle
 \nonumber\\
 &+\zeta
 |\Lambda=0,\Sigma= \tfrac12,\Omega=\tfrac12\rangle,
 \end{align}
where
 \begin{align}\label{so-mixing1}
 \zeta\sim A/(E_\Pi-E_\Sigma),
 \end{align}
 and then
 \begin{align}\label{Pi_1/2}
 \langle\Omega=\tfrac12|H_\mathrm{Cor}|\Omega=-\tfrac12\rangle
 =2\zeta B (J+\tfrac12) \langle\Lambda=1|L_x|\Lambda=0\rangle.
 \end{align}
Note that $\zeta$ depends on the non-diagonal matrix element of
spin-orbit interaction and \Eref{so-mixing1} is only an order of
magnitude estimate. It is important, though, that non-diagonal and
diagonal matrix elements have similar dependence on FC. We conclude
that $\Omega$-splitting for the $\Pi_{1/2}$ level must scale as
$ABJ/(E_\Pi-E_\Sigma)$.

The $\Omega$-doubling for $\Pi_{3/2}$ state takes place in the third
order in Coriolis interaction. Here $H_\mathrm{Cor}$ has to mix
first states $\Pi_{3/2}$ with $\Pi_{1/2}$ and $\Pi_{-3/2}$ with
$\Pi_{-1/2}$ before matrix element \eqref{Pi_1/2} can be used.
Therefore, the splitting scales as $B^3J^3/[A(E_\Pi-E_\Sigma)]$.

The above consideration corresponds to the coupling case $a$, when
$|A|\gg B$. In the opposite limit the states $\Pi_{1/2}$ and
$\Pi_{3/2}$ are strongly mixed by the Coriolis interaction and spin
$\bm{S}$ decouples from the molecular axis (coupling case $b$). As a
result, the quantum numbers $\Sigma$ and $\Omega$ are not defined
and we only have one quantum number $\Lambda=\pm 1$. Now
$\Lambda$-splitting takes place in the second order in Coriolis
interaction via intermediate $\Sigma$ state. The scaling here is
obviously of the form $B^2J^2/(E_\Pi-E_\Sigma)$. Note that in
contrast to the previous case $|A|\gg B$, the splitting here is
independent on $A$.

We can now use found scalings of the $\Lambda$- and
$\Omega$-doublings to determine sensitivity coefficients
\eqref{K-factors}. For this we only need to know how parameters $A$
and $B$ depend on $\alpha$ and $\beta$. In atomic units these
parameters obviously scale as: $A\propto\alpha^2$ and
$B\propto\beta$. We conclude, that for the case $a$ the
$\Omega$-doubling spectrum has following sensitivity coefficients:
 \begin{subequations}\label{doubling}
 \begin{align}
 \label{doubling1}
 &\mathrm{State\,}^2\Pi_{1/2}:
 \quad K_\alpha = \phantom{-}2\,,
 \quad K_\beta = 1\,,\\
 \label{doubling2}
 &\mathrm{State\,}^2\Pi_{3/2}:
 \quad K_\alpha = -2\,,
 \quad K_\beta = 3\,.
 \end{align}
For the case $b$, when $\bm{S}$ is completely decoupled from the
axis, the $\Lambda$-doubling spectrum has following sensitivity
coefficients:
 \begin{align}
 \label{doubling3}
 &\mathrm{State\,\,}\Pi:
 \quad K_\alpha = 0\,,
 \quad K_\beta = 2\,.
 \end{align}
 \end{subequations}

When constant $A$ is slightly larger than $B$, the spin $\bm{S}$ is
coupled to the axis only for lower rotational levels. As rotational
energy grows with $J$ and becomes larger than the splitting between
states $\Pi_{1/2}$ and $\Pi_{3/2}$, the spin decouples from the
axis. Consequently, the $\Omega$-doubling is transformed into
$\Lambda$-doubling. Equations \eqref{doubling} show that this can
cause significant changes in sensitivity coefficients. The
spin-orbit constant $A$ can be either positive (CH molecule), or
negative (OH). The sign of the $\Omega$-doubling depends on the sign
of $A$, while $\Lambda$-doubling does not depend on $A$ at all.
Therefore, decoupling of the spin can change the sign of the
splitting. In \sref{numeric} we will see that this can lead to the
dramatic enhancement of the sensitivity to the variation of FC.

\begin{figure*}[tbh]
 \includegraphics[scale=0.85]{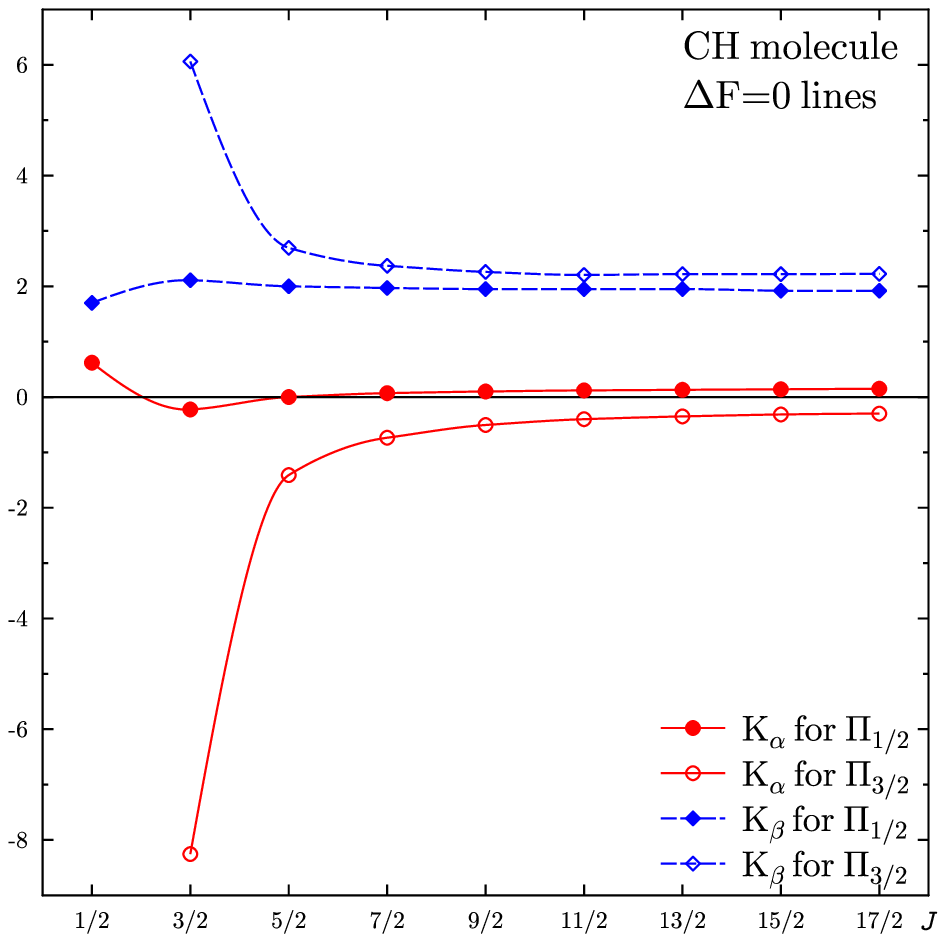}
 \hfill
 \includegraphics[scale=0.85]{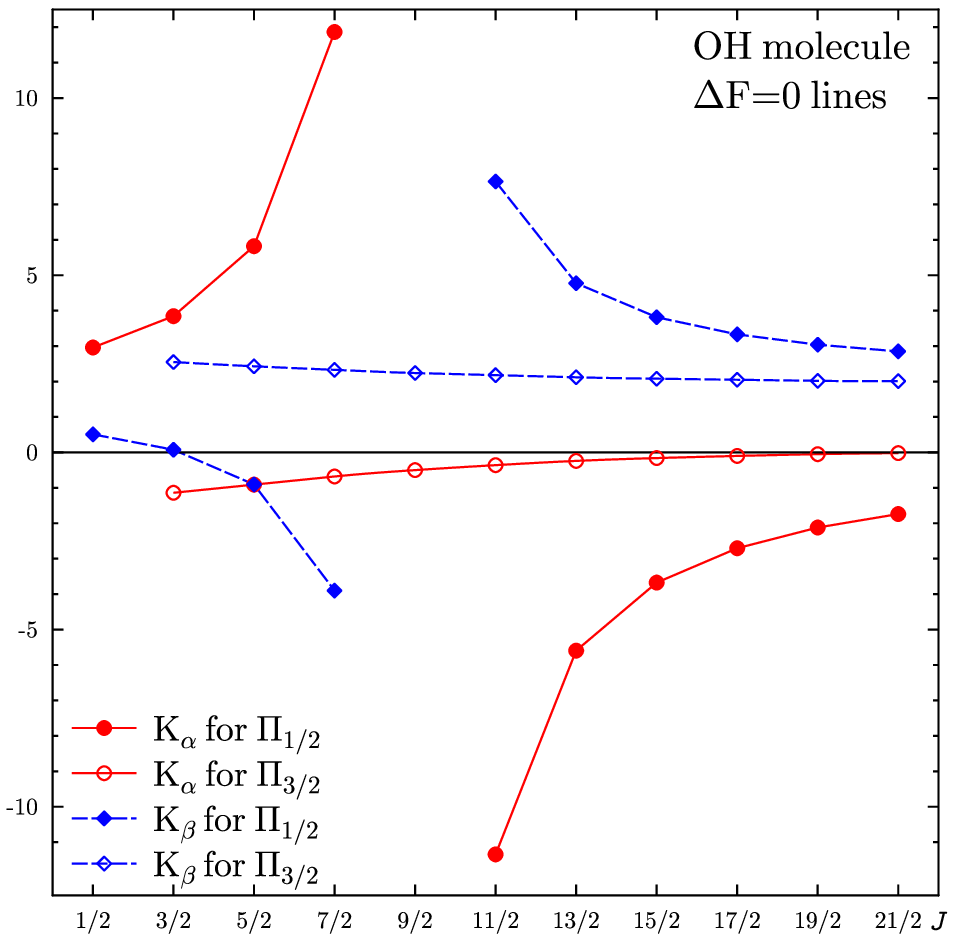}
 \caption{Sensitivity coefficients $K_\alpha$ and $K_\beta$ for
 $\Lambda$-doublet lines with
 $\Delta F=0$ in CH and OH molecules. The difference between lines
 with $F=J+\tfrac12$ and $F=J-\tfrac12$ is too small to be seen.
 For the state $\Pi_{3/2}$ of OH molecule the values for $J=\tfrac92$
 are too large to be shown on the plot. They are listed in \tref{tab_CH_OH}.}
 \label{fig_CH_OH}
\end{figure*}

\subsection{Intermediate coupling}\label{numeric}

$\Lambda$-doubling for the intermediate coupling was studied in
detail in many papers, including \cite{MD72,BCW79,BM79} (see also
the book \cite{BC03}). Here we use effective Hamiltonian
$H_\mathrm{eff}$ from Ref.\ \cite{MD72} in the subspace of the
levels $\Pi_{1/2}^\pm$ and $\Pi_{3/2}^\pm$, where upper sign
corresponds to parity $p$ in \Eref{patity_states}. Operator
$H_\mathrm{eff}$ includes spin-rotational and hyperfine
parts\footnote{Here we use notation $H_\mathrm{sr}$ to
define part of the effective Hamiltonian, which describes rotational
degrees of freedom and electron spin.}:
 \begin{align}\label{Heff1}
 H_\mathrm{eff} &=H_\mathrm{sr} + H_\mathrm{hf}\,.
 \end{align}
Neglecting third order terms in Coriolis and spin-orbit
interactions, we get the following simplified form of
spin-rotational part:
 \begin{subequations}\label{Heff2}
 \begin{align}
 \langle\Pi_{1/2}^\pm | H_\mathrm{sr} |\Pi_{1/2}^\pm\rangle
 &=-\tfrac12 A + B(J+\tfrac12)^2
 \nonumber\\
 &\pm(S_1+S_2)(2J+1)\,,
 \label{Heff2a}\\
 \langle\Pi_{3/2}^\pm | H_\mathrm{sr} |\Pi_{3/2}^\pm\rangle
 &=+\tfrac12 A + B(J+\tfrac12)^2 -2B\,,
 \label{Heff2b}\\
 \langle\Pi_{3/2}^\pm | H_\mathrm{sr} |\Pi_{1/2}^\pm\rangle
 &=\left[B \pm S_2
 (J+\tfrac12)\right]
 \nonumber\\
 &\times\sqrt{(J-\tfrac12)(J+\tfrac32)}\,,
 \label{Heff2c}
 \end{align}
 \end{subequations}
Here in addition to parameters $A$ and $B$ we have two parameters,
which appear in the second order of perturbation theory via
intermediate state(s) $\Sigma_{1/2}$. Parameter $S_1$ corresponds to
the cross term of the perturbation theory in spin-orbit and Coriolis
interactions, while parameter $S_2$ is quadratic in Coriolis
interaction. Because of this $S_1$ scales as $\alpha^2\beta$ and
$S_2$ scales as $\beta^2$. The third order parameters neglected in
\eqref{Heff2} consist of several terms each with different
dependencies on parameters $\alpha$ and $\beta$ \cite{MD72}. For
this reason we can not use them to study sensitivity coefficients.
Fortunately, all third order terms are very small for the molecules
considered here. They account only for the fine tuning of the
spectrum and do not noticeably affect sensitivity coefficients for
transitions with moderate quantum numbers $J$. It is easy to see
that Hamiltonian $H_\mathrm{sr}$ describes limiting cases $|A|\gg B$
and $|A|\ll B$ considered in Sec.\ \ref{analytic}.

The hyperfine part of effective Hamiltonian is defined in the lowest
order of perturbation theory and has the form:
 \begin{subequations}\label{Heff3}
 \begin{align}
 \langle\Pi_{1/2}^\pm | H_\mathrm{hf} |\Pi_{1/2}^\pm\rangle
 &= C_F\left[2a-b-c \pm (2J+1)d\right],
 \label{Heff3a}\\
 \langle\Pi_{3/2}^\pm | H_\mathrm{hf} |\Pi_{3/2}^\pm\rangle
 &=3C_F\left[2a+b+c\right],
 \label{Heff3b}\\
 \langle\Pi_{3/2}^\pm | H_\mathrm{hf} |\Pi_{1/2}^\pm\rangle
 &=-C_F\sqrt{(2J-1)(2J+3)}\,b\,,
 \label{Heff3c}
 \\
 C_F&\equiv\frac{F(F+1)-J(J+1)-I(I+1)}{8J(J+1)}\,.
 \nonumber
 \end{align}
 \end{subequations}
Here we assume that only one nucleus has spin and include only
magnetic dipole hyperfine interaction. In this approximation all
four parameters of $H_\mathrm{hf}$ scale as $\alpha^2\beta
g_\mathrm{nuc}$.

Effective Hamiltonian described by Eqs.\ (\ref{Heff2},\ref{Heff3})
has 8 parameters. We use NIST values \cite{NIST_diatomics} for the
fine structure splitting $A$, rotational constant $B$, and magnetic
hyperfine constants $a$, $b$, $c$, $d$. Remaining two parameters
$S_1$ and $S_2$ are found by minimizing \textit{rms} deviation
between theoretical and experimental $\Lambda$-doubling spectra.

In order to find sensitivity coefficients $K_\alpha$ we calculate
transition frequency for two values of $\alpha=\alpha_0\pm\delta$
near its physical value $\alpha_0=1/137.035999679(94)$ and similarly
for $K_\beta$ and $K_g$. We use scaling rules discussed above to
recalculate parameters of the effective Hamiltonian for different
values of FC. Then we use numerical differentiation to find
respective sensitivity coefficient.

We check the accuracy of our approach by adding three most important
third order parameters from Ref.\ \cite{MD72} to Hamiltonian
\eqref{Heff2} and including them in fitting procedure. That leads to
noticeable improvement of the theoretical frequencies for higher
values of $J$. Each of our three third order parameters actually
include several terms, which scale as different combination of $A$
and $B$ ($A^2B$, $AB^2$, etc.) Each term, therefore, has different
dependence on $\alpha$ and $\beta$. On the other hand, they have
same dependence on the quantum numbers and can not be independently
determined from the fitting procedure. Because of that it is
impossible to unambiguously determine dependence of these parameters
on FC. Therefore, we calculate sensitivity coefficients assuming
dominance of one term for each third order parameter and look how
the answer depends on these assumptions. We found that sensitivity
coefficients changed by less than 1\%. Therefore, we conclude that
this simple model is sufficiently accurate for our purposes and
currently there is no need to use more elaborate theory.

Hyperfine Hamiltonian \eqref{Heff3} accounts only for one nuclear
spin and does not include interaction with nuclear electric
quadrupole moment. Generalization to two spins is straightforward,
but in this paper we restrict consideration to molecules with one
spin. For molecules with $I>\tfrac12$
we must add quadrupole term to Eqs.\ \eqref{Heff3}:
 \begin{align}
 \langle\Pi_\Omega^\pm | \tilde{H}_\mathrm{hf} |\Pi_\Omega^\pm\rangle
 &=
 \frac{C(C+1)-4I(I+1)J(J+1)}
      {8I(2I-1)J(J+1)(2J+3)}
 \label{Heff4}\\
 &\times  (-1)^{2J}
 \left[3\Omega^2-J(J+1)\right]
 \left(eq_0 Q_\mathrm{nuc}\right),
 \nonumber\\
 C&\equiv F(F+1)-J(J+1)-I(I+1)\,.
 \nonumber
 \end{align}

In this case there is additional hyperfine parameter $eq_0
Q_\mathrm{nuc}$ which includes electronic matrix element $eq_0$ and
nuclear quadrupole moment $Q_\mathrm{nuc}$. Matrix element $eq_0$
for light molecules can be calculated in non-relativistic
approximation and does not depend on FC.
Dependence of $Q_\mathrm{nuc}$ on FC can be very complex (see
discussion in \cite{FW09}). Without going into nuclear theory, one
can consider $Q_\mathrm{nuc}$ as independent fundamental parameter
and introduce additional sensitivity coefficient $K_Q$. Below we
will see that coefficients $K_g$ and $K_Q$ are usually very small,
except for the transitions with very low frequency.

\section{Results and discussion}\label{discussion}

\begin{table*}[bht!]
\caption{Frequencies (in MHz) and sensitivity coefficients for
hyperfine components $(J,F\rightarrow J,F')$ of $\Lambda$-doublet
lines in CH and OH molecules. Recommended frequencies and their
uncertainties are taken from Refs.~\cite{NIST_diatomics,JPL_Catalog,CDMS}.}

\label{tab_CH_OH}

\begin{tabular}{lldccdcddddd}
 \hline\hline\\[-7pt]
 \multicolumn{1}{c}{Molecule}
 &\multicolumn{1}{c}{Level}
 &\multicolumn{1}{c}{$J$} &\multicolumn{1}{c}{$F$}
 &\multicolumn{1}{c}{$F'$} &\multicolumn{4}{c}{$\omega$~(MHz)}
 &\multicolumn{1}{c}{$K_\alpha$}
 &\multicolumn{1}{c}{$K_\beta$}
 &\multicolumn{1}{c}{$K_g$}\\
 \cline{6-9}\\[-7pt]
&&&&&\multicolumn{1}{c}{Recom.} &\multicolumn{1}{c}{Uncert.}
&\multicolumn{1}{c}{Theory} &\multicolumn{1}{c}{Diff.}
\\
\hline
 $^{12}$CH & $^2\Pi_{1/2}$
 & 0.5 & 0 & 1 &   3263.795&   0.003 & 3269.40 &  -5.61 &  0.59  &   1.71 &  -0.02  \\
&& 0.5 & 1 & 1 &   3335.481&   0.001 & 3340.77 &  -5.29 &  0.62  &   1.70 &   0.00  \\
&& 0.5 & 1 & 0 &   3349.194&   0.003 & 3354.11 &  -4.92 &  0.63  &   1.69 &   0.01  \\[1mm]
&& 1.5 & 1 & 2 &   7275.004&   0.001 & 7262.25 &  12.75 & -0.24  &   2.12 &  -0.01  \\
&& 1.5 & 1 & 1 &   7325.203&   0.001 & 7312.02 &  13.18 & -0.23  &   2.11 &   0.00  \\
&& 1.5 & 2 & 2 &   7348.419&   0.001 & 7335.30 &  13.12 & -0.22  &   2.11 &   0.00  \\
&& 1.5 & 2 & 1 &   7398.618&   0.001 & 7385.08 &  13.54 & -0.20  &   2.10 &   0.01  \\[1mm]
 $^{12}$CH & $^2\Pi_{3/2}$
 & 1.5 & 2 & 2 &    701.68 &   0.01  &  682.96 &  18.72 & -8.44  &   6.15 &  -0.01  \\
&& 1.5 & 1 & 2 &    703.97 &   0.03  &  679.83 &  24.14 & -8.66  &   6.32 &  -0.01  \\
&& 1.5 & 2 & 1 &    722.30 &   0.03  &  702.98 &  19.52 & -8.37  &   6.17 &   0.02  \\
&& 1.5 & 1 & 1 &    724.79 &   0.01  &  699.85 &  24.94 & -8.07  &   5.97 &   0.02  \\[1mm]
 $^{16}$OH & $^2\Pi_{3/2}$
 & 1.5 & 1 & 2 &   1612.2310& 0.0002 & 1595.42 &  16.81 &  -1.27 &   2.61 &  -0.03  \\
&& 1.5 & 1 & 1 &   1665.4018& 0.0002 & 1648.93 &  16.47 &  -1.14 &   2.55 &   0.00  \\
&& 1.5 & 2 & 2 &   1667.3590& 0.0002 & 1650.66 &  16.70 &  -1.14 &   2.55 &   0.00  \\
&& 1.5 & 2 & 1 &   1720.5300& 0.0002 & 1704.17 &  16.36 &  -1.02 &   2.49 &   0.03  \\[1mm]
 $^{16}$OH & $^2\Pi_{1/2}$
 & 0.5 & 0 & 1 &   4660.2420& 0.0030 & 4638.98 &  21.26 &   2.98 &   0.50 &  -0.02  \\
&& 0.5 & 1 & 1 &   4750.6560& 0.0030 & 4729.51 &  21.15 &   2.96 &   0.51 &   0.00  \\
&& 0.5 & 1 & 0 &   4765.5620& 0.0030 & 4744.50 &  21.06 &   2.96 &   0.51 &   0.01  \\[1mm]
&& 4.5 & 5 & 4 &     88.9504& 0.0011 &   64.34 &  24.61 &-921.58 & 459.86 &  -0.56  \\
&& 4.5 & 5 & 5 &    117.1495& 0.0011 &   92.35 &  24.80 &-699.65 & 349.59 &  -0.19  \\
&& 4.5 & 4 & 4 &    164.7960& 0.0011 &  141.20 &  23.60 &-496.67 & 248.77 &   0.16  \\
&& 4.5 & 4 & 5 &    192.9957& 0.0011 &  169.22 &  23.78 &-424.05 & 212.68 &   0.28  \\
\hline\hline
\end{tabular}
\end{table*}

We applied the above method to $^{16}$OH, $^{12}$CH,
$^{7}$Li$^{16}$O, $^{14}$N$^{16}$O, and $^{15}$N$^{16}$O. Molecules
CH and NO have ground state $^2\Pi_{1/2}$ ($A>0$), while OH and LiO
have ground state $^2\Pi_{3/2}$ ($A<0$). The ratio $|A/B|$ changes
from 2 for CH molecule, to 7 for OH, and to almost a hundred for LiO
and NO. Therefore, LiO and NO definitely belong to the coupling case
$a$. For OH molecule we can expect transition from case $a$ for
lower rotational states to case $b$ for higher ones. Finally, for CH
we expect intermediate coupling for lower rotational states and
coupling case $b$ for higher states.

Let us see how this scheme works in practice for the effective
Hamiltonian (\ref{Heff2},\ref{Heff3}). Fig.\ \ref{fig_CH_OH}
demonstrate $J$-dependence of sensitivity coefficients for CH and OH
molecules. Both of them have only one nuclear spin $I=\tfrac12$. For
a given quantum number $J$, each $\Lambda$-doublet transition has
four hyperfine components: two strong transitions with $\Delta F=0$
and $F=J\pm\tfrac12$ (for $J=\tfrac12$ there is only one transition
with $F=1$) and two weaker transitions with $\Delta F=\pm 1$. The
hyperfine structure for OH and CH molecules is rather small and
sensitivity coefficients for all hyperfine components are very
close. Because of that Fig.\ \ref{fig_CH_OH} presents only averaged
values for strong transitions with $\Delta F=0$.

We see that for large values of $J$ the sensitivity coefficients for
both molecules approach limit \eqref{doubling3} of the coupling case
$b$. The opposite limits (\ref{doubling1},\ref{doubling2}) are not
reached for either molecule even for smallest values of $J$. So, we
conclude that coupling case $a$ is not realized. It is interesting
that in Fig.\ \ref{fig_CH_OH} the curves for the lower states are
smooth, while for upper states there are singularities. For CH
molecule this singularity takes place for the state $\Pi_{3/2}$ near
the lowest possible value $J=3/2$. Singularity for OH molecule takes
place for state $\Pi_{1/2}$ near $J=9/2$.

These singularities appear because $\Lambda$ splitting turns to
zero. As we saw above, the sign of the splitting for the coupling
case $a$ depends on the sign of the constant $A$. The same sign
determines which state $\Pi_{1/2}$, or $\Pi_{3/2}$ lies higher. As a
result, for the lower state the sign of the splitting is the same
for both limiting cases, but decoupling of the electron spin $S$ for
the upper state leads to the change of sign of the splitting. Of
course, these singularities are most interesting for our purposes,
as they lead to large sensitivity coefficients which strongly depend
on the quantum numbers. Note, that when the frequency of the
transition is small, it becomes sensitive to the hyperfine part of
the Hamiltonian and sensitivity coefficients for hyperfine
components may differ significantly. Sensitivity coefficients of all
hyperfine components of such $\Lambda$-lines are given in
\tref{tab_CH_OH}. We can see that near the singularities all
sensitivity coefficients, including $K_g$, are enhanced.

\begin{figure}[h!]
 \includegraphics[scale=0.85]{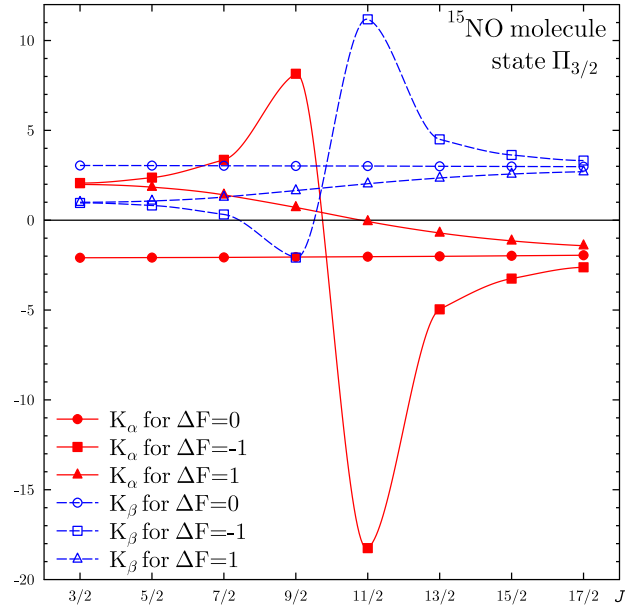}
 \caption{Sensitivity coefficients for $\Lambda$-doublet lines in $\Pi_{3/2}$
 state of $^{15}$NO molecule. The difference between two hyperfine components
 with $\Delta F=0$ is too small to be seen. Sensitivity coefficients for $\Pi_{1/2}$
 state correspond to the coupling case $a$ (see \Eref{doubling1}).}
 \label{fig_NO}
\end{figure}

Now let us consider sensitivity coefficients for the molecule
$^{15}$NO. Here we expect expressions for the coupling case $a$ to
be applicable. In fact, for the state $\Pi_{1/2}$ coefficients
$K_\alpha$ and $K_\beta$ agree with prediction \eqref{doubling1}
within few percent and $K_g\ll 1$. However, for the state
$\Pi_{3/2}$ \Eref{doubling2} works only for transitions with $\Delta
F=0$, see Fig.\ \ref{fig_NO}. Indeed, $\Lambda$-splitting for low
values of $J$ is smaller, than hyperfine structure. As a result, the
frequencies of $\Delta F=\pm 1$ transitions strongly depend on the
hyperfine parameters. For some values of $J$ these frequencies can
be very small, because $\Lambda$-splitting and hyperfine splitting
cancel each other. This leads to enhancement of sensitivity
coefficients, similar to that, discussed in Ref.\ \cite{Fla06b}.
Fig.\ \ref{fig_NO} shows that for $^{15}$NO molecule such
singularity takes place for $\Delta F=-1$ transition near
$J=\tfrac{11}{2}$. For smaller values of $J$ the hyperfine
contribution to transition frequency dominates over
$\Lambda$-splitting. Sensitivity coefficients for this case are
similar to those of the normal hyperfine transitions, i.e.
$K_\alpha\approx 2$ and $K_\beta\approx K_g\approx 1$. For higher
values of $J$ they approach the limit \eqref{doubling2}. For $\Delta
F=1$ transitions there is no singularity and sensitivities change
smoothly between same limiting values. Finally, the hyperfine energy
for the lines with $\Delta F=0$ is negligible and these lines are
described by \Eref{doubling2} for all values of $J$.

\begin{table*}[tb!]
\caption{Frequencies (in MHz) and sensitivity coefficients for
$\Lambda$-doublet lines in $^{14}$N$^{16}$O. Experimental
frequencies and their uncertainties are taken from
Refs.~\cite{NIST_diatomics,JPL_Catalog}.}

\label{tab_14N16O}

\begin{tabular}{ldccdddddddd}
\hline\hline\\[-7pt]
 \multicolumn{1}{c}{Level}
&\multicolumn{1}{c}{$J$} &\multicolumn{1}{c}{$F$}
&\multicolumn{1}{c}{$F'$} &\multicolumn{4}{c}{$\omega$~(MHz)}
&\multicolumn{1}{c}{$K_\alpha$} &\multicolumn{1}{c}{$K_\beta$}
&\multicolumn{1}{c}{$K_g$} &\multicolumn{1}{c}{$K_Q$}\\
\cline{5-8}\\[-7pt]
&&&&\multicolumn{1}{c}{Exper.} &\multicolumn{1}{c}{Uncert.}
&\multicolumn{1}{c}{Theory} &\multicolumn{1}{c}{Diff.}
\\
\hline
$^2\Pi_{1/2}$
& 0.5 & 0.5 & 0.5 &  205.9510&0.0002 &  205.96 &  -0.01 &  1.95  &   1.03 &  -0.73 &  0.00 \\
& 0.5 & 0.5 & 1.5 &  225.9357&0.0002 &  225.93 &   0.00 &  1.95  &   1.02 &  -0.58 &  0.00 \\
& 0.5 & 1.5 & 0.5 &  411.2056&0.0002 &  411.19 &   0.01 &  1.97  &   1.01 &   0.13 &  0.00 \\
& 0.5 & 1.5 & 1.5 &  431.1905&0.0002 &  431.16 &   0.03 &  1.97  &   1.01 &   0.17 &  0.00 \\[1mm]
& 1.5 & 0.5 & 0.5 &  560.8538&0.0002 &  561.22 &  -0.37 &  1.97  &   1.02 &  -0.27 &  0.00 \\
& 1.5 & 0.5 & 1.5 &  587.7467&0.0002 &  587.54 &   0.21 &  1.97  &   1.01 &  -0.21 &  0.09 \\
& 1.5 & 1.5 & 0.5 &  624.6494&0.0002 &  624.93 &  -0.28 &  1.97  &   1.02 &  -0.14 & -0.09 \\
& 1.5 & 1.5 & 1.5 &  651.5425&0.0002 &  651.25 &   0.29 &  1.97  &   1.01 &  -0.09 &  0.00 \\
& 1.5 & 1.5 & 2.5 &  693.8282&0.0002 &  693.88 &  -0.05 &  1.98  &   1.01 &  -0.02 & -0.04 \\
& 1.5 & 2.5 & 1.5 &  758.9106&0.0002 &  758.66 &   0.25 &  1.97  &   1.01 &   0.06 &  0.04 \\
& 1.5 & 2.5 & 2.5 &  801.1963&0.0002 &  801.29 &  -0.10 &  1.98  &   1.01 &   0.11 &  0.00 \\[1mm]
& 2.5 & 3.5 & 3.5 & 1160.7768&0.0003 & 1160.95 &  -0.18 &  1.98  &   1.01 &   0.08 &  0.00 \\
& 3.5 & 4.5 & 4.5 & 1514.768 &0.001  & 1514.97 &  -0.20 &  2.00  &   1.00 &   0.07 &  0.00 \\[1mm]
$^2\Pi_{3/2}$
& 1.5 & 0.5 & 0.5 &          &       &    0.81 &        & -2.09  &   3.05 &  -0.11 &  0.00 \\
& 1.5 & 1.5 & 1.5 &    0.612 &0.001  &    0.87 &  -0.25 & -2.96  &   4.31 &  -0.06 &  0.00 \\
& 1.5 & 2.5 & 2.5 &    1.029 &0.001  &    0.96 &   0.07 & -1.94  &   2.83 &   0.05 &  0.00 \\
& 1.5 & 0.5 & 1.5 &          &       &   44.45 &        &  2.09  &   0.93 &   1.01 &  1.25 \\
& 1.5 & 1.5 & 0.5 &   46.464 &0.003  &   46.12 &   0.34 &  1.92  &   1.00 &   0.96 &  1.19 \\
& 1.5 & 1.5 & 2.5 &   73.286 &0.003  &   73.33 &  -0.05 &  2.10  &   0.96 &   1.02 & -0.42 \\
& 1.5 & 2.5 & 1.5 &   74.931 &0.003  &   75.15 &  -0.22 &  2.00  &   1.02 &   1.00 & -0.41 \\[1mm]
& 2.5 & 1.5 & 1.5 &          &       &    3.38 &        & -2.08  &   3.04 &  -0.06 &  0.00 \\
& 2.5 & 2.5 & 2.5 &    3.121 &0.001  &    3.54 &  -0.41 & -2.36  &   3.44 &  -0.02 &  0.00 \\
& 2.5 & 3.5 & 3.5 &    3.923 &0.001  &    3.75 &   0.17 & -1.99  &   2.91 &   0.04 &  0.00 \\
& 2.5 & 1.5 & 2.5 &          &       &   27.37 &        &  2.64  &   0.70 &   1.14 & -0.97 \\
& 2.5 & 2.5 & 1.5 &   34.39  &0.03   &   34.28 &   0.11 &  1.68  &   1.17 &   0.90 & -0.77 \\
& 2.5 & 2.5 & 3.5 &   40.172 &       &   40.07 &   0.10 &  2.46  &   0.76 &   1.08 &  0.46 \\
& 2.5 & 3.5 & 2.5 &   47.211 &0.001  &   47.36 &  -0.14 &  1.77  &   1.11 &   0.92 &  0.39 \\[1mm]
& 3.5 & 2.5 & 2.5 &          &       &    8.58 &        & -2.07  &   3.03 &  -0.04 &  0.00 \\
& 3.5 & 3.5 & 3.5 &          &       &    8.88 &        & -2.07  &   3.03 &  -0.01 &  0.00 \\
& 3.5 & 4.5 & 4.5 &          &       &    9.26 &        & -2.07  &   3.03 &   0.03 &  0.00 \\
& 3.5 & 2.5 & 3.5 &          &       &   13.59 &        &  5.06  &  -0.39 &   1.73 & -6.83 \\
& 3.5 & 3.5 & 2.5 &   31.550 &0.004  &   31.05 &   0.50 &  1.03  &   1.51 &   0.73 & -2.94 \\
& 3.5 & 3.5 & 4.5 &          &       &   21.54 &        &  3.90  &  -0.01 &   1.38 &  3.32 \\
& 3.5 & 4.5 & 3.5 &   39.221 &0.002  &   39.67 &  -0.45 &  1.18  &   1.40 &   0.76 &  1.82 \\[1mm]
& 4.5 & 3.5 & 3.5 &          &       &   17.25 &        & -2.05  &   3.02 &  -0.03 &  0.00 \\
& 4.5 & 4.5 & 4.5 &          &       &   17.74 &        & -2.05  &   3.02 &  -0.01 &  0.00 \\
& 4.5 & 5.5 & 5.5 &          &       &   18.33 &        & -2.05  &   3.03 &   0.03 &  0.00 \\
& 4.5 & 3.5 & 4.5 &          &       &    0.96 &        &-80.19  &  38.95 & -19.07 &157.32 \\
& 4.5 & 4.5 & 3.5 &   35.045 &0.002  &   34.02 &   1.02 &  0.16  &   1.95 &   0.50 & -4.33 \\
& 4.5 & 4.5 & 5.5 &          &       &    5.26 &        & 16.68  &  -6.81 &   4.16 & 23.52 \\
& 4.5 & 5.5 & 4.5 &   40.512 &0.001  &   41.33 &  -0.81 &  0.34  &   1.81 &   0.55 &  3.05 \\
\hline\hline
\end{tabular}
\end{table*}

\begin{table*}[htb]
\caption{Frequencies (in MHz) and sensitivity coefficients for
$\Lambda$-doublet lines in $^{7}$Li$^{16}$O. Experimental
frequencies and their uncertainties are taken from
Ref.~\cite{NIST_diatomics}.}

\label{tab_7Li16O}

\begin{tabular}{ldccdddddddd}
\hline\hline\\[-7pt]
 \multicolumn{1}{c}{Level}
&\multicolumn{1}{c}{$J$} &\multicolumn{1}{c}{$F$}
&\multicolumn{1}{c}{$F'$} &\multicolumn{4}{c}{$\omega$~(MHz)}
&\multicolumn{1}{c}{$K_\alpha$} &\multicolumn{1}{c}{$K_\beta$}
&\multicolumn{1}{c}{$K_g$} &\multicolumn{1}{c}{$K_Q$}\\
\cline{5-8}\\[-7pt]
&&&&\multicolumn{1}{c}{Exper.} &\multicolumn{1}{c}{Uncert.}
&\multicolumn{1}{c}{Theory} &\multicolumn{1}{c}{Diff.}
\\
\hline
$^2\Pi_{3/2}$
& 1.5 & 1 & 1 &    11.28 & 0.01  &    11.18 &   0.10 & -1.90 &  2.94 &  0.00 &  0.00  \\
& 1.5 & 2 & 2 &    11.28 & 0.01  &    11.18 &   0.10 & -1.90 &  2.94 &  0.00 &  0.00  \\
& 1.5 & 3 & 3 &    11.28 & 0.01  &    11.19 &   0.09 & -1.90 &  2.94 &  0.00 &  0.00  \\
& 1.5 & 0 & 1 &          &       &     9.82 &        & -2.51 &  3.25 & -0.14 &  0.63  \\
& 1.5 & 1 & 0 &          &       &    12.53 &        & -1.46 &  2.74 &  0.11 & -0.49  \\
& 1.5 & 1 & 2 &     8.55 & 0.10  &     8.41 &   0.14 & -3.26 &  3.59 & -0.33 &  0.72  \\
& 1.5 & 2 & 1 &    14.00 & 0.05  &    13.95 &   0.05 & -1.07 &  2.55 &  0.20 & -0.44  \\
& 1.5 & 2 & 3 &     6.95 & 0.05  &     6.87 &   0.08 & -4.47 &  4.24 & -0.61 & -0.89  \\
& 1.5 & 3 & 2 &    15.60 & 0.02  &    15.50 &   0.10 & -0.76 &  2.36 &  0.27 &  0.40  \\[1mm]
& 2.5 & 1 & 1 &    45.02 & 0.03  &    44.80 &   0.22 & -1.90 &  2.95 &  0.00 &  0.00  \\
& 2.5 & 2 & 2 &    45.02 & 0.03  &    44.80 &   0.22 & -1.90 &  2.95 &  0.00 &  0.00  \\
& 2.5 & 3 & 3 &    45.02 & 0.03  &    44.81 &   0.21 & -1.90 &  2.95 &  0.00 &  0.00  \\
& 2.5 & 4 & 4 &    45.02 & 0.03  &    44.83 &   0.19 & -1.90 &  2.95 &  0.00 &  0.00  \\
& 2.5 & 1 & 2 &    44.04 & 0.03  &    43.82 &   0.22 & -2.01 &  3.01 & -0.02 & -0.08  \\
& 2.5 & 2 & 1 &    45.97 & 0.03  &    45.78 &   0.19 & -1.81 &  2.90 &  0.02 &  0.08  \\
& 2.5 & 2 & 3 &    43.60 & 0.03  &    43.37 &   0.23 & -2.06 &  3.03 & -0.03 & -0.05  \\
& 2.5 & 3 & 2 &    46.43 & 0.03  &    46.24 &   0.19 & -1.76 &  2.88 &  0.03 &  0.04  \\
& 2.5 & 3 & 4 &    43.17 & 0.03  &    42.97 &   0.20 & -2.11 &  3.06 & -0.04 &  0.08  \\
& 2.5 & 4 & 3 &    46.86 & 0.03  &    46.67 &   0.19 & -1.71 &  2.86 &  0.04 & -0.08  \\[1mm]
%
& 3.5 & 3 & 3 &   112.237& 0.002 &   112.28 &  -0.04 & -1.91 &  2.96 &  0.00 &  0.00  \\
& 4.5 & 4 & 4 &   223.756& 0.003 &   225.24 &  -1.48 & -1.91 &  2.99 &  0.00 &  0.00  \\[1mm]
$^2\Pi_{1/2}$
& 0.5 & 1 & 1 &          &       &  2958.44 &        &  2.10 &  0.95 &  0.00 &  0.00  \\
& 0.5 & 2 & 2 &          &       &  2969.20 &        &  2.10 &  0.95 &  0.00 &  0.00  \\
& 0.5 & 1 & 2 &          &       &  2947.37 &        &  2.10 &  0.95 & -0.01 &  0.00  \\
& 0.5 & 2 & 1 &          &       &  2980.27 &        &  2.10 &  0.95 &  0.01 &  0.00  \\[1mm]
& 1.5 & 1 & 1 &          &       &  5971.29 &        &  2.10 &  0.96 &  0.00 &  0.00  \\
& 1.5 & 2 & 2 &          &       &  5975.59 &        &  2.10 &  0.96 &  0.00 &  0.00  \\
& 1.5 & 3 & 3 &          &       &  5982.05 &        &  2.10 &  0.96 &  0.00 &  0.00  \\
& 1.5 & 1 & 2 &          &       &  5969.71 &        &  2.10 &  0.96 &  0.00 &  0.00  \\
& 1.5 & 2 & 1 &          &       &  5977.18 &        &  2.10 &  0.96 &  0.00 &  0.00  \\
& 1.5 & 2 & 3 &          &       &  5973.38 &        &  2.10 &  0.96 &  0.00 &  0.00  \\
& 1.5 & 3 & 2 &          &       &  5984.26 &        &  2.10 &  0.96 &  0.00 &  0.00  \\[1mm]
%
& 2.5 & 2 & 2 &          &       &  9078.46 &        &  2.11 &  0.97 &  0.00 &  0.00  \\[1mm]
& 3.5 & 2 & 2 & 12321.51 & 0.03  & 12322.14 &  -0.63 &  2.13 &  0.99 &  0.00 &  0.00  \\
& 3.5 & 3 & 3 & 12325.82 & 0.03  & 12325.21 &   0.61 &  2.13 &  0.99 &  0.00 &  0.00  \\
& 3.5 & 2 & 3 & 12319.93 & 0.03  & 12321.60 &  -1.67 &  2.13 &  0.99 &  0.00 &  0.00  \\
& 3.5 & 3 & 2 & 12327.41 & 0.03  & 12325.75 &   1.66 &  2.13 &  0.99 &  0.00 &  0.00  \\
\hline\hline
\end{tabular}
\end{table*}

The spectrum and sensitivity coefficients of the molecule $^{14}$NO
are similar to those of $^{15}$NO. Because $^{14}$N has nuclear spin
$I=1$, the hyperfine structure of the $\Lambda$-doublet lines is
more complex and consists of up to 7 hyperfine components. Hyperfine
Hamiltonian includes magnetic dipole part \eqref{Heff3} and electric
quadrupole part \eqref{Heff4} and is described by five hyperfine
parameters, which we take from Ref.\ \cite{NIST_diatomics}. As we
said above, we are not discussing nuclear theory here and consider
nuclear quadrupole moment as independent FC. Because of that
$\Lambda$-doublet spectrum is now described by four sensitivity
coefficients (see \tref{tab_14N16O}).

Sensitivity coefficients $K_\alpha$ and $K_\beta$ of the
$\Lambda$-doublet lines of the state $\Pi_{1/2}$ again agree with
\eqref{doubling1} within few percent. The lowest frequency
transitions for $J=\tfrac12$ have sensitivity coefficients $K_g$ of
the order of unity, but they rapidly decrease with frequency and
with $J$. Coefficients $K_Q$ for the state $\Pi_{1/2}$ are always
small.

For the state $\Pi_{3/2}$ there are transitions of three types.
First type transitions correspond to $\Delta F=0$. The hyperfine
energy difference here is small compared to $\Lambda$-splitting.
These transitions have sensitivity coefficients $K_\alpha$ and
$K_\beta$ close to prediction \eqref{doubling2} and small
coefficients $K_g$ and $K_Q$. Second type transitions correspond to
$\Delta F=\pm 1$ and small values of $J$. Hyperfine energy for these
transitions dominates over $\Lambda$-splitting. Sensitivity
coefficients here are close to those of pure hyperfine transitions,
i.e. $K_\alpha=2$ and $K_\beta=1$. As long as hyperfine energy
includes comparable magnetic dipole and electric quadrupole parts,
coefficients $K_g$ and $K_Q$ are of the order of unity, but may
significantly differ from one transition to another. Note that all
transitions of this type for $^{15}$NO molecule have $K_g \approx
1$.

Transitions of the third type also correspond to $\Delta F=\pm 1$,
but higher rotational quantum numbers $J=\tfrac72,\,\tfrac92$. The
hyperfine transition energy here is comparable to
$\Lambda$-splitting and they can either double, or almost cancel
each other. Consequently, sensitivity coefficient are widely spread
and can become very large for transitions with anomalously low
frequency.

Note that low frequency transitions for $J=\tfrac92$ were not
observed experimentally and we use theoretical frequencies to
calculate sensitivity coefficients. Because of significant
cancelation of different contributions, the accuracy of these
frequencies can be low. When these frequencies are measured,
respective sensitivity coefficients should be corrected:
\begin{align}\label{correction}
 K_{i,\mathrm{cor}}=K_i
 \frac{\omega_\mathrm{theor}}{\omega_\mathrm{exper}}
\end{align}

Sensitivity coefficients for LiO molecule are listed in
\tref{tab_7Li16O}. The hyperfine structure here is smaller than for
NO molecule and sensitivity coefficients are closer to case $a$
values (\ref{doubling1},\ref{doubling2}). Significant deviations are
found only for $J=\tfrac32,\,\Delta F =\pm 1$ transitions of
$\Pi_{3/2}$ state. Also, these are the only transitions, where
coefficients $K_g$ and $K_Q$ are not negligible. For this molecule
there are no transitions with anomalously small frequencies and,
therefore, sensitivity coefficients are not enhanced.

\section{Conclusions}

In this paper we calculated sensitivity coefficients to variation of
fundamental constants for $\Lambda$-doublet spectra of several light
diatomic molecules. We found several lines with anomalously high
sensitivity. All these lines have relatively low frequencies and
enhanced sensitivity is caused by the significant cancelations
between contributions from different parts of the effective
Hamiltonian \eqref{Heff1}.

In CH and OH molecules enhancement takes place when electron spin
decouples from the molecular axis and $\Omega$-doubling is
transformed into $\Lambda$-doubling. For one of the two states
$\Pi_{1/2}$, or $\Pi_{3/2}$ this transformation leads to the change
of sign of the splitting between states with definite parity and
enhanced sensitivity to FC variation.

Rotational constant $B$ for $^{14}$NO and $^{15}$NO molecules is
much smaller, than for CH and NH molecules and electron spin is
strongly coupled to the axis. Consequently, there is no enhancement
caused by decoupling. On the other hand, the hyperfine structure of
the $\Lambda$-doublet lines is comparable to $\Lambda$-splitting in
$\Pi_{3/2}$ state. For some transition lines with $\Delta F=\pm 1$
the hyperfine energy almost cancel $\Lambda$-splitting leading to
enhanced sensitivity.

For LiO molecule electron spin is strongly coupled to the axis and
hyperfine structure is smaller than $\Lambda$-splitting. As a
result, there is no strong enhancement of the sensitivity to FC
variation. However, even here sensitivity coefficients strongly
depend on the quantum numbers.

Sensitivity coefficients for $\Lambda$-doublet transitions of OH
molecules were calculated before in Refs.\ \cite{KCL05,KC04}. For
all these states our results are in good agreement with those
calculations. In particular, from \tref{tab_CH_OH} we find
sensitivity coefficients for the 18 cm ground state
$\Lambda$-doublet transitions with $J=\tfrac32$ and $F'=F$ to be:
$K_\alpha=-1.14$, $K_\beta=2.55$, and $K_g= 0$. In the the paper
\cite{KCL05} the 21-cm Hydrogen line was used as a reference. It has
$K_\alpha=2$, $K_\beta=1$, and $K_g=1$. Parameter $\cal F$ according
to \Eref{redshifts2} is given by the expression:
 \begin{align}\label{tildeF2}
  {\cal F}=\alpha^{\Delta K_\alpha}\beta^{\Delta K_\beta}
  g_\mathrm{nuc}^{\Delta K_g}
  =\alpha^{3.14}\beta^{-1.55}g_\mathrm{nuc}^{1}\,.
 \end{align}
This result is sufficiently close to \Eref{tildeF}.

For astrophysical observations it is important to have accurate
laboratory measurements so that frequency ratios for distant object
can be compared to the respective local ratios. Sufficiently
accurate frequency measurements were done only for 18 cm lines of OH
\cite{LMH06,HLS06} and for 9 cm lines of CH \cite{MMB06}. These
lines can be used for new studies of the variation of FC without
significant preliminary work. For other lines at present there are
no sufficiently accurate laboratory frequencies. New laboratory
measurements are necessary before these lines can be used for our
purposes. In particular, the hyperfine components of the 42 cm CH
line are most interesting as they have high sensitivity to both
fundamental constants and were already observed in astrophysics for
distant objects.

In principle it is possible to study time variation of FC without
referring to the laboratory measurements. For this purpose it is
possible to compare microwave spectra for molecular clouds from our
Galaxy with extragalactic spectra of the same species. In many cases
the line widths for the galactic spectra are one-two orders of
magnitude smaller, than for objects at cosmological distances, so
they can serve as very good reference.

Let us briefly discuss the feasibility of the laboratory tests of
time-variation of FC using molecular $\Lambda$-doublets. Present
model independent laboratory limit on $\beta$-variation is
\cite{SBC08}:
 \begin{align}\label{lab_beta}
 \frac{\mathrm{d}\beta/\mathrm{d}t}{\beta}=(3.8\pm
 5.6)10^{-14}\,\mathrm{yr}^{-1}\,,
 \end{align}
and the limit on $\alpha$-variation is three orders of magnitude
stronger, on the level $10^{-17}$ \cite{Ros08}. To improve
constrained \eqref{lab_beta} one needs to measure frequency shifts
$\delta\omega < K_\beta \omega \delta\beta/\beta$. For the 18 cm OH
line this corresponds to the shift $\delta\omega \lesssim 4\times
10^{-4}$~Hz. This is few orders of magnitude smaller than the
accuracy of the best present measurements \cite{LMH06,HLS06}. On the
other hand, at present there is rapid progress in precision
molecular spectroscopy caused by development of sources of ultracold
molecules (see review \cite{CDKY09} and references therein). Thus it
is possible that molecular tests of FC variation using
$\Lambda$-doublet lines can become competitive in the near future.

When comparing sensitivity of  different laboratory experiments on
time-variation it is not sufficient to look for large dimensionless
sensitivity coefficients \eqref{K-factors}. In high precision
laboratory measurements the line widths are not dominated by the
Doppler effect and are not proportional to the frequency. Because of
that, instead of the dimensionless sensitivity coefficients $K_i$,
which determine relative frequency shifts \eqref{K-factors}, one has
to look for large absolute sensitivities $K_i\omega$, which
determine absolute frequency shifts $\delta\omega$ and for narrow
lines. In astrophysics, on the contrary, all lines are
Doppler-broadened and dimensionless sensitivity coefficients $K_i$
become crucial.


\begin{acknowledgments}
The author is grateful to Christian Henkel and Karl Menten for
attracting his attention to this problem, to Sergey Porsev for help
with numerical estimates of sensitivity coefficients, to Sergey
Levshakov for interesting discussions, and to the referee for
constructive criticism. This research is partly supported by RFBR
grants 08-02-00460 and 09-02-12223.
\end{acknowledgments}


\begin{thebibliography}{40}
\expandafter\ifx\csname
natexlab\endcsname\relax\def\natexlab#1{#1}\fi
\expandafter\ifx\csname bibnamefont\endcsname\relax
  \def\bibnamefont#1{#1}\fi
\expandafter\ifx\csname bibfnamefont\endcsname\relax
  \def\bibfnamefont#1{#1}\fi
\expandafter\ifx\csname citenamefont\endcsname\relax
  \def\citenamefont#1{#1}\fi
\expandafter\ifx\csname url\endcsname\relax
  \def\url#1{\texttt{#1}}\fi
\expandafter\ifx\csname urlprefix\endcsname\relax\def\urlprefix{URL
}\fi \providecommand{\bibinfo}[2]{#2}
\providecommand{\eprint}[2][]{\url{#2}}

\bibitem[{\citenamefont{Shelkovnikov et~al.}(2008)\citenamefont{Shelkovnikov,
  Butcher, Chardonnet, and {Amy-Klein}}}]{SBC08}
\bibinfo{author}{\bibfnamefont{A.}~\bibnamefont{Shelkovnikov}},
  \bibinfo{author}{\bibfnamefont{R.~J.} \bibnamefont{Butcher}},
  \bibinfo{author}{\bibfnamefont{C.}~\bibnamefont{Chardonnet}},
  \bibnamefont{and}
  \bibinfo{author}{\bibfnamefont{A.}~\bibnamefont{{Amy-Klein}}},
  \bibinfo{journal}{Phys. Rev. Lett.} \textbf{\bibinfo{volume}{100}},
  \bibinfo{pages}{150801} (\bibinfo{year}{2008}),
  \eprint{arXiv:\eprint{0803.1829}}.

\bibitem[{\citenamefont{Rosenband et~al.}(2008)}]{Ros08}
\bibinfo{author}{\bibfnamefont{T.}~\bibnamefont{Rosenband}}
  \bibnamefont{et~al.}, \bibinfo{journal}{Science}
  \textbf{\bibinfo{volume}{319}}, \bibinfo{pages}{1808} (\bibinfo{year}{2008}).

\bibitem[{\citenamefont{Murphy et~al.}(2003)\citenamefont{Murphy, Webb, and
  Flambaum}}]{MWF03}
\bibinfo{author}{\bibfnamefont{M.~T.} \bibnamefont{Murphy}},
  \bibinfo{author}{\bibfnamefont{J.~K.} \bibnamefont{Webb}}, \bibnamefont{and}
  \bibinfo{author}{\bibfnamefont{V.~V.} \bibnamefont{Flambaum}},
  \bibinfo{journal}{Mon. Not. R. Astron. Soc.} \textbf{\bibinfo{volume}{345}},
  \bibinfo{pages}{609} (\bibinfo{year}{2003}),
  \bibinfo{note}{arXiv:\eprint{astro-ph/0310318}}.

\bibitem[{\citenamefont{Srianand et~al.}(2004)\citenamefont{Srianand, Chand,
  Petitjean, and Aracil}}]{SCP04}
\bibinfo{author}{\bibfnamefont{R.}~\bibnamefont{Srianand}},
  \bibinfo{author}{\bibfnamefont{H.}~\bibnamefont{Chand}},
  \bibinfo{author}{\bibfnamefont{P.}~\bibnamefont{Petitjean}},
  \bibnamefont{and} \bibinfo{author}{\bibfnamefont{B.}~\bibnamefont{Aracil}},
  \bibinfo{journal}{Phys. Rev. Lett.} \textbf{\bibinfo{volume}{92}},
  \bibinfo{pages}{121302} (\bibinfo{year}{2004}),
  \eprint{arXiv:astro-ph/0402177}.

\bibitem[{\citenamefont{Levshakov et~al.}(2007)\citenamefont{Levshakov, Molaro,
  Lopez et~al.}}]{LML07}
\bibinfo{author}{\bibfnamefont{S.~A.} \bibnamefont{Levshakov}},
  \bibinfo{author}{\bibfnamefont{P.}~\bibnamefont{Molaro}},
  \bibinfo{author}{\bibfnamefont{S.}~\bibnamefont{Lopez}},
  \bibnamefont{et~al.}, \bibinfo{journal}{Astron.~Astrophys.}
  \textbf{\bibinfo{volume}{466}}, \bibinfo{pages}{1077} (\bibinfo{year}{2007}),
  \eprint{arXiv:astro-ph/0703042}.

\bibitem[{\citenamefont{Shlyakhter}(1976)}]{Shl76}
\bibinfo{author}{\bibfnamefont{A.~I.} \bibnamefont{Shlyakhter}},
  \bibinfo{journal}{Nature} \textbf{\bibinfo{volume}{264}},
  \bibinfo{pages}{340} (\bibinfo{year}{1976}).

\bibitem[{\citenamefont{Flambaum and Wiringa}(2009)}]{FW09}
\bibinfo{author}{\bibfnamefont{V.~V.} \bibnamefont{Flambaum}} \bibnamefont{and}
  \bibinfo{author}{\bibfnamefont{R.~B.} \bibnamefont{Wiringa}},
  \bibinfo{journal}{Phys. Rev. C} \textbf{\bibinfo{volume}{79}},
  \bibinfo{pages}{034302} (\bibinfo{year}{2009}), \eprint{arXiv:0807.4943}.

\bibitem[{\citenamefont{{Essen} et~al.}(1971)\citenamefont{{Essen},
  {Donaldson}, {Bangham}, and {Hope}}}]{EDB71}
\bibinfo{author}{\bibfnamefont{L.}~\bibnamefont{{Essen}}},
  \bibinfo{author}{\bibfnamefont{R.~W.} \bibnamefont{{Donaldson}}},
  \bibinfo{author}{\bibfnamefont{M.~J.} \bibnamefont{{Bangham}}},
  \bibnamefont{and} \bibinfo{author}{\bibfnamefont{E.~G.}
  \bibnamefont{{Hope}}}, \bibinfo{journal}{Nature}
  \textbf{\bibinfo{volume}{229}}, \bibinfo{pages}{110} (\bibinfo{year}{1971}).

\bibitem[{\citenamefont{Hudson et~al.}(2006)\citenamefont{Hudson, Lewandowski,
  Sawyer, and Ye}}]{HLS06}
\bibinfo{author}{\bibfnamefont{E.~R.} \bibnamefont{Hudson}},
  \bibinfo{author}{\bibfnamefont{H.~J.} \bibnamefont{Lewandowski}},
  \bibinfo{author}{\bibfnamefont{B.~C.} \bibnamefont{Sawyer}},
  \bibnamefont{and} \bibinfo{author}{\bibfnamefont{J.}~\bibnamefont{Ye}},
  \bibinfo{journal}{Phys. Rev. Lett.} \textbf{\bibinfo{volume}{96}},
  \bibinfo{pages}{143004} (\bibinfo{year}{2006}).

\bibitem[{\citenamefont{{McCarthy} et~al.}(2006)\citenamefont{{McCarthy},
  {Mohamed}, {Brown}, and {Thaddeus}}}]{MMB06}
\bibinfo{author}{\bibfnamefont{M.~C.} \bibnamefont{{McCarthy}}},
  \bibinfo{author}{\bibfnamefont{S.}~\bibnamefont{{Mohamed}}},
  \bibinfo{author}{\bibfnamefont{J.~M.} \bibnamefont{{Brown}}},
  \bibnamefont{and}
  \bibinfo{author}{\bibfnamefont{P.}~\bibnamefont{{Thaddeus}}},
  \bibinfo{journal}{Proc. of the Nat. Academy of Science}
  \textbf{\bibinfo{volume}{103}}, \bibinfo{pages}{12263}
  (\bibinfo{year}{2006}).

\bibitem[{\citenamefont{Ziurys and Turner}(1985)}]{ZT85}
\bibinfo{author}{\bibfnamefont{L.~M.} \bibnamefont{Ziurys}} \bibnamefont{and}
  \bibinfo{author}{\bibfnamefont{B.~E.} \bibnamefont{Turner}},
  \bibinfo{journal}{Astrophys. J.} \textbf{\bibinfo{volume}{292}},
  \bibinfo{pages}{L25} (\bibinfo{year}{1985}).

\bibitem[{\citenamefont{Kukolich}(1967)}]{Kuk67}
\bibinfo{author}{\bibfnamefont{S.~G.} \bibnamefont{Kukolich}},
  \bibinfo{journal}{Phys. Rev.} \textbf{\bibinfo{volume}{156}},
  \bibinfo{pages}{83} (\bibinfo{year}{1967}).

\bibitem[{\citenamefont{{Savedoff}}(1956)}]{Sav56}
\bibinfo{author}{\bibfnamefont{M.~P.} \bibnamefont{{Savedoff}}},
  \bibinfo{journal}{Nature} \textbf{\bibinfo{volume}{178}},
  \bibinfo{pages}{688} (\bibinfo{year}{1956}).

\bibitem[{\citenamefont{Darling}(2003)}]{Dar03}
\bibinfo{author}{\bibfnamefont{J.}~\bibnamefont{Darling}},
  \bibinfo{journal}{Phys. Rev. Lett.} \textbf{\bibinfo{volume}{91}},
  \bibinfo{pages}{011301} (\bibinfo{year}{2003}).

\bibitem[{\citenamefont{{Chengalur} and {Kanekar}}(2003)}]{CK03}
\bibinfo{author}{\bibfnamefont{J.~N.} \bibnamefont{{Chengalur}}}
  \bibnamefont{and}
  \bibinfo{author}{\bibfnamefont{N.}~\bibnamefont{{Kanekar}}},
  \bibinfo{journal}{Phys. Rev. Lett.} \textbf{\bibinfo{volume}{91}},
  \bibinfo{pages}{241302} (\bibinfo{year}{2003}),
  \eprint{arXiv:astro-ph/0310764}.

\bibitem[{\citenamefont{{Kanekar} et~al.}(2004)\citenamefont{{Kanekar},
  {Chengalur}, and {Ghosh}}}]{KCG04}
\bibinfo{author}{\bibfnamefont{N.}~\bibnamefont{{Kanekar}}},
  \bibinfo{author}{\bibfnamefont{J.~N.} \bibnamefont{{Chengalur}}},
  \bibnamefont{and} \bibinfo{author}{\bibfnamefont{T.}~\bibnamefont{{Ghosh}}},
  \bibinfo{journal}{Phys. Rev. Lett.} \textbf{\bibinfo{volume}{93}},
  \bibinfo{pages}{051302} (\bibinfo{year}{2004}),
  \eprint{arXiv:astro-ph/0406121}.

\bibitem[{\citenamefont{{van Veldhoven} et~al.}(2004)\citenamefont{{van
  Veldhoven}, K\"{u}pper, Bethlem, Sartakov, {van Roij}, and Meijer}}]{VKB04}
\bibinfo{author}{\bibfnamefont{J.}~\bibnamefont{{van Veldhoven}}},
  \bibinfo{author}{\bibfnamefont{J.}~\bibnamefont{K\"{u}pper}},
  \bibinfo{author}{\bibfnamefont{H.~L.} \bibnamefont{Bethlem}},
  \bibinfo{author}{\bibfnamefont{B.}~\bibnamefont{Sartakov}},
  \bibinfo{author}{\bibfnamefont{A.~J.~A.} \bibnamefont{{van Roij}}},
  \bibnamefont{and} \bibinfo{author}{\bibfnamefont{G.}~\bibnamefont{Meijer}},
  \bibinfo{journal}{Eur. Phys. J. D} \textbf{\bibinfo{volume}{31}},
  \bibinfo{pages}{337} (\bibinfo{year}{2004}).

\bibitem[{\citenamefont{Flambaum and Kozlov}(2007)}]{FK07a}
\bibinfo{author}{\bibfnamefont{V.~V.} \bibnamefont{Flambaum}} \bibnamefont{and}
  \bibinfo{author}{\bibfnamefont{M.~G.} \bibnamefont{Kozlov}},
  \bibinfo{journal}{Phys. Rev. Lett.} \textbf{\bibinfo{volume}{98}},
  \bibinfo{pages}{240801} (\bibinfo{year}{2007}), \eprint{arXiv: 0704.2301}.

\bibitem[{\citenamefont{Murphy et~al.}(2001)\citenamefont{Murphy, Webb,
  Flambaum, Drinkwater, Combes, and Wiklind}}]{MWF01c}
\bibinfo{author}{\bibfnamefont{M.~T.} \bibnamefont{Murphy}},
  \bibinfo{author}{\bibfnamefont{J.~K.} \bibnamefont{Webb}},
  \bibinfo{author}{\bibfnamefont{V.~V.} \bibnamefont{Flambaum}},
  \bibinfo{author}{\bibfnamefont{M.~J.} \bibnamefont{Drinkwater}},
  \bibinfo{author}{\bibfnamefont{F.}~\bibnamefont{Combes}}, \bibnamefont{and}
  \bibinfo{author}{\bibfnamefont{T.}~\bibnamefont{Wiklind}},
  \bibinfo{journal}{Mon. Not. R. Astron. Soc.} \textbf{\bibinfo{volume}{327}},
  \bibinfo{pages}{1244} (\bibinfo{year}{2001}).

\bibitem[{\citenamefont{Kanekar et~al.}(2005)\citenamefont{Kanekar, Carilli,
  Langston et~al.}}]{KCL05}
\bibinfo{author}{\bibfnamefont{N.}~\bibnamefont{Kanekar}},
  \bibinfo{author}{\bibfnamefont{C.~L.} \bibnamefont{Carilli}},
  \bibinfo{author}{\bibfnamefont{G.~I.} \bibnamefont{Langston}},
  \bibnamefont{et~al.}, \bibinfo{journal}{Phys. Rev. Lett.}
  \textbf{\bibinfo{volume}{95}}, \bibinfo{pages}{261301}
  (\bibinfo{year}{2005}).

\bibitem[{\citenamefont{Murphy et~al.}(2008)\citenamefont{Murphy, Flambaum,
  Muller, and Henkel}}]{MFMH08}
\bibinfo{author}{\bibfnamefont{M.~T.} \bibnamefont{Murphy}},
  \bibinfo{author}{\bibfnamefont{V.~V.} \bibnamefont{Flambaum}},
  \bibinfo{author}{\bibfnamefont{S.}~\bibnamefont{Muller}}, \bibnamefont{and}
  \bibinfo{author}{\bibfnamefont{C.}~\bibnamefont{Henkel}},
  \bibinfo{journal}{Science} \textbf{\bibinfo{volume}{320}},
  \bibinfo{pages}{1611} (\bibinfo{year}{2008}),
  \eprint{\eprint{arXiv:0806.3081}}.

\bibitem[{\citenamefont{Levshakov
  et~al.}(2008{\natexlab{a}})\citenamefont{Levshakov, Reimers, Kozlov, Porsev,
  and Molaro}}]{LRK08}
\bibinfo{author}{\bibfnamefont{S.~A.} \bibnamefont{Levshakov}},
  \bibinfo{author}{\bibfnamefont{D.}~\bibnamefont{Reimers}},
  \bibinfo{author}{\bibfnamefont{M.~G.} \bibnamefont{Kozlov}},
  \bibinfo{author}{\bibfnamefont{S.~G.} \bibnamefont{Porsev}},
  \bibnamefont{and} \bibinfo{author}{\bibfnamefont{P.}~\bibnamefont{Molaro}},
  \bibinfo{journal}{Astron.~Astrophys.} \textbf{\bibinfo{volume}{479}},
  \bibinfo{pages}{719} (\bibinfo{year}{2008}{\natexlab{a}}), \eprint{arXiv:
  \eprint{0712.2890}}.

\bibitem[{\citenamefont{Kanekar and Chengalur}(2004)}]{KC04}
\bibinfo{author}{\bibfnamefont{N.}~\bibnamefont{Kanekar}} \bibnamefont{and}
  \bibinfo{author}{\bibfnamefont{J.~N.} \bibnamefont{Chengalur}},
  \bibinfo{journal}{Mon. Not. R. Astron. Soc.} \textbf{\bibinfo{volume}{350}},
  \bibinfo{pages}{L17} (\bibinfo{year}{2004}), \eprint{arXiv:astro-ph/0310765}.

\bibitem[{\citenamefont{Rydbeck et~al.}(1974)\citenamefont{Rydbeck, Ellder,
  Sume, Hjalmarson, and Irvine}}]{RES74}
\bibinfo{author}{\bibfnamefont{O.~E.~H.} \bibnamefont{Rydbeck}},
  \bibinfo{author}{\bibfnamefont{J.}~\bibnamefont{Ellder}},
  \bibinfo{author}{\bibfnamefont{A.}~\bibnamefont{Sume}},
  \bibinfo{author}{\bibfnamefont{A.}~\bibnamefont{Hjalmarson}},
  \bibnamefont{and} \bibinfo{author}{\bibfnamefont{W.~M.}
  \bibnamefont{Irvine}}, \bibinfo{journal}{Astron.~Astrophys.}
  \textbf{\bibinfo{volume}{34}}, \bibinfo{pages}{479} (\bibinfo{year}{1974}).

\bibitem[{\citenamefont{Henkel and Menten}(2008)}]{HM08}
\bibinfo{author}{\bibfnamefont{C.}~\bibnamefont{Henkel}} \bibnamefont{and}
  \bibinfo{author}{\bibfnamefont{K.}~\bibnamefont{Menten}}
  (\bibinfo{year}{2008}), \bibinfo{note}{private communication}.

\bibitem[{\citenamefont{Mart{\'{\i}}n et~al.}(2003)\citenamefont{Mart{\'{\i}}n,
  Mauersberger, {Mart{\'{\i}}n-Pintado}, {Garc{\'{\i}}a-Burillo}, and
  Henkel}}]{MMM03}
\bibinfo{author}{\bibfnamefont{S.}~\bibnamefont{Mart{\'{\i}}n}},
  \bibinfo{author}{\bibfnamefont{R.}~\bibnamefont{Mauersberger}},
  \bibinfo{author}{\bibfnamefont{J.}~\bibnamefont{{Mart{\'{\i}}n-Pintado}}},
  \bibinfo{author}{\bibfnamefont{S.}~\bibnamefont{{Garc{\'{\i}}a-Burillo}}},
  \bibnamefont{and} \bibinfo{author}{\bibfnamefont{C.}~\bibnamefont{Henkel}},
  \bibinfo{journal}{Astron.~Astrophys.} \textbf{\bibinfo{volume}{411}},
  \bibinfo{pages}{L465} (\bibinfo{year}{2003}),
  \eprint{arXiv:\eprint{astro-ph/0309663}}.

\bibitem[{\citenamefont{Pagani et~al.}(2009)\citenamefont{Pagani, Daniel, and
  Dubernet}}]{PDD09}
\bibinfo{author}{\bibfnamefont{L.}~\bibnamefont{Pagani}},
  \bibinfo{author}{\bibfnamefont{F.}~\bibnamefont{Daniel}}, \bibnamefont{and}
  \bibinfo{author}{\bibfnamefont{M.-L.} \bibnamefont{Dubernet}},
  \bibinfo{journal}{Astron.~Astrophys.} \textbf{\bibinfo{volume}{494}},
  \bibinfo{pages}{719} (\bibinfo{year}{2009}), \eprint{arXiv:
  \eprint{0811.3289}}.

\bibitem[{\citenamefont{{Lev} et~al.}(2006)\citenamefont{{Lev}, {Meyer},
  {Hudson}, {Sawyer}, {Bohn}, and {Ye}}}]{LMH06}
\bibinfo{author}{\bibfnamefont{B.~L.} \bibnamefont{{Lev}}},
  \bibinfo{author}{\bibfnamefont{E.~R.} \bibnamefont{{Meyer}}},
  \bibinfo{author}{\bibfnamefont{E.~R.} \bibnamefont{{Hudson}}},
  \bibinfo{author}{\bibfnamefont{B.~C.} \bibnamefont{{Sawyer}}},
  \bibinfo{author}{\bibfnamefont{J.~L.} \bibnamefont{{Bohn}}},
  \bibnamefont{and} \bibinfo{author}{\bibfnamefont{J.}~\bibnamefont{{Ye}}},
  \bibinfo{journal}{Phys. Rev. A} \textbf{\bibinfo{volume}{74}},
  \bibinfo{pages}{061402(R)} (\bibinfo{year}{2006}),
  \eprint{arXiv:physics/0608194}.

\bibitem[{\citenamefont{Levshakov
  et~al.}(2008{\natexlab{b}})\citenamefont{Levshakov, Molaro, and
  Kozlov}}]{LMK08}
\bibinfo{author}{\bibfnamefont{S.~A.} \bibnamefont{Levshakov}},
  \bibinfo{author}{\bibfnamefont{P.}~\bibnamefont{Molaro}}, \bibnamefont{and}
  \bibinfo{author}{\bibfnamefont{M.~G.} \bibnamefont{Kozlov}}
  (\bibinfo{year}{2008}{\natexlab{b}}), \bibinfo{note}{arXiv:{0808.0583}}.

\bibitem[{\citenamefont{{Henkel} et~al.}(2009)\citenamefont{{Henkel}, {Menten},
  {Murphy}, {Jethava}, {Flambaum}, {Braatz}, {Muller}, {Ott}, and
  {Mao}}}]{KMM09}
\bibinfo{author}{\bibfnamefont{C.}~\bibnamefont{{Henkel}}},
  \bibinfo{author}{\bibfnamefont{K.~M.} \bibnamefont{{Menten}}},
  \bibinfo{author}{\bibfnamefont{M.~T.} \bibnamefont{{Murphy}}},
  \bibinfo{author}{\bibfnamefont{N.}~\bibnamefont{{Jethava}}},
  \bibinfo{author}{\bibfnamefont{V.~V.} \bibnamefont{{Flambaum}}},
  \bibinfo{author}{\bibfnamefont{J.~A.} \bibnamefont{{Braatz}}},
  \bibinfo{author}{\bibfnamefont{S.}~\bibnamefont{{Muller}}},
  \bibinfo{author}{\bibfnamefont{J.}~\bibnamefont{{Ott}}}, \bibnamefont{and}
  \bibinfo{author}{\bibfnamefont{R.~Q.} \bibnamefont{{Mao}}},
  \bibinfo{journal}{Astron.~Astrophys.} \textbf{\bibinfo{volume}{500}},
  \bibinfo{pages}{725} (\bibinfo{year}{2009}),
  \eprint{arXiv:\eprint{0904.3081}}.

\bibitem[{\citenamefont{{Carr} et~al.}(2009)\citenamefont{{Carr}, {DeMille},
  {Krems}, and {Ye}}}]{CDKY09}
\bibinfo{author}{\bibfnamefont{L.~D.} \bibnamefont{{Carr}}},
  \bibinfo{author}{\bibfnamefont{D.}~\bibnamefont{{DeMille}}},
  \bibinfo{author}{\bibfnamefont{R.~V.} \bibnamefont{{Krems}}},
  \bibnamefont{and} \bibinfo{author}{\bibfnamefont{J.}~\bibnamefont{{Ye}}},
  \bibinfo{journal}{New Journal of Physics} \textbf{\bibinfo{volume}{11}},
  \bibinfo{pages}{055049} (\bibinfo{year}{2009}), \eprint{arXiv:0904.3175}.

\bibitem[{\citenamefont{Lovas et~al.}()}]{NIST_diatomics}
\bibinfo{author}{\bibfnamefont{F.~J.} \bibnamefont{Lovas}}
  \bibnamefont{et~al.}, \emph{\bibinfo{title}{{Diatomic Spectral Database}}},
  \eprint{\url{http://www.physics.nist.gov/PhysRefData/MolSpec/Diatomic/index.%
html}}.

\bibitem[{\citenamefont{Chen et~al.}()\citenamefont{Chen, Cohen, Crawford,
  Drouin, Pearson, and Pickett}}]{JPL_Catalog}
\bibinfo{author}{\bibfnamefont{P.}~\bibnamefont{Chen}},
  \bibinfo{author}{\bibfnamefont{E.~A.} \bibnamefont{Cohen}},
  \bibinfo{author}{\bibfnamefont{T.~J.} \bibnamefont{Crawford}},
  \bibinfo{author}{\bibfnamefont{B.~J.} \bibnamefont{Drouin}},
  \bibinfo{author}{\bibfnamefont{J.~C.} \bibnamefont{Pearson}},
  \bibnamefont{and} \bibinfo{author}{\bibfnamefont{H.~M.}
  \bibnamefont{Pickett}}, \emph{\bibinfo{title}{{JPL Molecular Spectroscopy
  Catalog}}}, \eprint{\url{http://spec.jpl.nasa.gov/}}.

\bibitem[{\citenamefont{M\"{u}ller et~al.}()\citenamefont{M\"{u}ller,
  Schl\"{o}der, Stutzki, and Winnewisser}}]{CDMS}
\bibinfo{author}{\bibfnamefont{H.~S.~P.} \bibnamefont{M\"{u}ller}},
  \bibinfo{author}{\bibfnamefont{F.}~\bibnamefont{Schl\"{o}der}},
  \bibinfo{author}{\bibfnamefont{J.}~\bibnamefont{Stutzki}}, \bibnamefont{and}
  \bibinfo{author}{\bibfnamefont{G.}~\bibnamefont{Winnewisser}},
  \emph{\bibinfo{title}{{The Cologne Database for Molecular Spectroscopy
  (CDMS)}}}, \eprint{\url{http://www.astro.uni-koeln.de/site/vorhersagen/}}.

\bibitem[{\citenamefont{Griest et~al.}(2009)\citenamefont{Griest, Whitmore,
  Wolfe, Prochaska, Howk, and Marcy}}]{GWW09}
\bibinfo{author}{\bibfnamefont{K.}~\bibnamefont{Griest}},
  \bibinfo{author}{\bibfnamefont{J.~B.} \bibnamefont{Whitmore}},
  \bibinfo{author}{\bibfnamefont{A.~M.} \bibnamefont{Wolfe}},
  \bibinfo{author}{\bibfnamefont{J.~X.} \bibnamefont{Prochaska}},
  \bibinfo{author}{\bibfnamefont{J.~C.} \bibnamefont{Howk}}, \bibnamefont{and}
  \bibinfo{author}{\bibfnamefont{G.~W.} \bibnamefont{Marcy}}
  (\bibinfo{year}{2009}), \bibinfo{note}{arXiv:\eprint{0904.4725}}.

\bibitem[{\citenamefont{Meerts and Dymanus}(1972)}]{MD72}
\bibinfo{author}{\bibfnamefont{W.~L.} \bibnamefont{Meerts}} \bibnamefont{and}
  \bibinfo{author}{\bibfnamefont{A.}~\bibnamefont{Dymanus}},
  \bibinfo{journal}{J. Molecular Spectroscopy} \textbf{\bibinfo{volume}{44}},
  \bibinfo{pages}{320} (\bibinfo{year}{1972}).

\bibitem[{\citenamefont{Brown and Carrington}(2003)}]{BC03}
\bibinfo{author}{\bibfnamefont{J.}~\bibnamefont{Brown}} \bibnamefont{and}
  \bibinfo{author}{\bibfnamefont{A.}~\bibnamefont{Carrington}},
  \emph{\bibinfo{title}{Rotational Spectroscopy of Diatomic Molecules}}
  (\bibinfo{publisher}{Cambridge University Press}, \bibinfo{year}{2003}), ISBN
  \bibinfo{isbn}{0521810094}.

\bibitem[{\citenamefont{Brown et~al.}(1979)\citenamefont{Brown, Colbourn,
  Watson, and Wayne}}]{BCW79}
\bibinfo{author}{\bibfnamefont{J.~M.} \bibnamefont{Brown}},
  \bibinfo{author}{\bibfnamefont{E.~A.} \bibnamefont{Colbourn}},
  \bibinfo{author}{\bibfnamefont{J.~K.~G.} \bibnamefont{Watson}},
  \bibnamefont{and} \bibinfo{author}{\bibfnamefont{F.~D.} \bibnamefont{Wayne}},
  \bibinfo{journal}{J. of Molecular Spectroscopy}
  \textbf{\bibinfo{volume}{74}}, \bibinfo{pages}{294} (\bibinfo{year}{1979}).

\bibitem[{\citenamefont{Brown and Merer}(1979)}]{BM79}
\bibinfo{author}{\bibfnamefont{J.~M.} \bibnamefont{Brown}} \bibnamefont{and}
  \bibinfo{author}{\bibfnamefont{A.~J.} \bibnamefont{Merer}},
  \bibinfo{journal}{J. of Molecular Spectroscopy}
  \textbf{\bibinfo{volume}{74}}, \bibinfo{pages}{488} (\bibinfo{year}{1979}).

\bibitem[{\citenamefont{Flambaum}(2006)}]{Fla06b}
\bibinfo{author}{\bibfnamefont{V.~V.} \bibnamefont{Flambaum}},
  \bibinfo{journal}{Phys. Rev. A} \textbf{\bibinfo{volume}{73}},
  \bibinfo{pages}{034101} (\bibinfo{year}{2006}).

\end{thebibliography}

\end{document}